\documentclass[12pt]{article}

\usepackage{enumitem} 
\usepackage[utf8]{inputenc} 
\usepackage{amsmath, amsthm, amssymb, amsfonts,mathtools,physics,bbm, authblk}
\numberwithin{equation}{section}
\usepackage{subcaption}
\usepackage[multiple]{footmisc}
\usepackage{float}
\usepackage[textsize=scriptsize]{todonotes}
\usepackage[
    a4paper,
    margin=1in,
]{geometry}

\usepackage[linktoc=page]{hyperref}
\usepackage{cleveref}
\allowdisplaybreaks

\usepackage{tikz}
\usepackage{graphicx}
\graphicspath{{./fig}}

\usepackage[
    backend=bibtex,
    bibstyle=numeric,
    citestyle=numeric-comp,
    maxbibnames=99,
    giveninits=true,
    sorting=none,
]{biblatex}

\bibliography{LA2DR_paper.bbl}

\newbibmacro{string+doi}[1]{%
\iffieldundef{doi}{#1}{\href{http://dx.doi.org/\thefield{doi}}{#1}}}

\DeclareFieldFormat{title}{\usebibmacro{string+doi}{\mkbibemph{#1}}}
\DeclareFieldFormat[article]{title}{\usebibmacro{string+doi}{\mkbibemph{#1}}}
\DeclareFieldFormat[article,periodical]{journaltitle}{{#1}} 
\DeclareFieldFormat[article]{pages}{{#1}} 
\DeclareFieldFormat[article]{volume}{\mkbibbold{#1}} 

\DeclareFieldFormat[book]{title}{\usebibmacro{string+doi}{{\sl #1}}}

\DeclareFieldFormat[incollection]{title}{\usebibmacro{string+doi}{{\em #1}}}

\renewbibmacro{in:}{%
\ifentrytype{article}
{}
{\printtext{\bibstring{in}\intitlepunct}}}

\AtEveryBibitem{\clearfield{isbn}}
\AtEveryBibitem{\clearfield{editor}}

\def\d{\partial}

\def\cp{\mathbb{CP}}
\def\bar{\overline}

\def\bphi{\overline{\varphi}}
\def\tphi{\tilde{\varphi}}
\def\bpsi{\overline{\psi}}

\def\sgn{\operatorname{sgn}}
\def\srb{S[\varphi_{\rm rb}]}
\def\srbz{S_{\rm rb}^{(0)}}

\renewcommand*{\tilde}[1]{\widetilde{#1}}

\def\vp{\varphi}
\def\phib{\bar{\varphi}}
\def\tphi{\tilde{\vp}}
\def\db{\bar{\d}}
\def\psib{\bar{\psi}}
\def\chib{\bar{\chi}}
\def\Db{\bar{D}}
\def\zb{\bar{z}}
\def\xb#1{x_{B #1}}

\def\ab#1{\alpha_{B #1}}
\def\xib{\bar{\xi}}
\def\Gammab{\bar{\Gamma}}
\def\seff{S_{\rm eff}}

\DeclareMathOperator{\arccosh}{arccosh}
\DeclareMathOperator{\arctanh}{arctanh}

\def\RR{\mathbb{R}}

\def\ZZ{\mathbb{Z}}

\def\cB{\mathcal{B}}
\def\cC{\mathcal{C}}
\def\cD{\mathcal{D}}

\def\cG{\mathcal{G}}
\def\cH{\mathcal{H}}
\def\cI{\mathcal{I}}
\def\cJ{\mathcal{J}}
\def\cK{\mathcal{K}}
\def\cL{\mathcal{L}}
\def\cM{\mathcal{M}}
\def\cN{\mathcal{N}}

\def\cR{\mathcal{R}}

\title{\bf\Large Bions and infrared renormalons\\[2mm] in K\"ahler sigma models with $U(1)$ isotropy \vspace{5mm}}
\author[]{Chao-Hsiang Sheu
}
\affil[]{\small\it  
	Quantum Universe Center, Korea Institute for Advanced Study, Seoul 02455, Korea}
\date{}

\begin{document}
\noindent
\begin{flushright}
KIAS-Q26034\\ 
\today
\end{flushright}

\begingroup
\let\newpage\relax
\maketitle
\endgroup

\begin{abstract}
We study semiclassical bion contributions to the vacuum energy of K\"ahler sigma models with two-dimensional target spaces on $\RR\times S^1$ with twisted boundary conditions.
This is first illustrated in the deformed $\cp^1$ model, where explicit real and complex bion solutions are constructed in a manner analogous to that used in its quantum-mechanical reduction \cite{Sheu:2024hae}.
Evaluating the bosonic and fermionic fluctuations at one loop, we obtain the effective action with both the coupling constant and the elongation parameter properly renormalized. Integration over the bion quasi-moduli space using Lefschetz thimbles then gives the leading nonperturbative contribution, which the real and complex bion contributions cancel at the supersymmetric point, while the first imaginary ambiguity appears when deviating from the supersymmetric point. Requiring its cancellation against the perturbative ambiguity allows us to infer the corresponding infrared-renormalon pole.
This is an explicit example generalizing the resurgent structure of the $\mathbb{CP}^1$ model in \cite{Fujimori:2018kqp} from one coupling to two renormalized parameters.
The calculation further extends to K\"ahler targets of one complex dimension with a $U(1)$ symmetry.
At leading dilute one-loop order, the target dependence enters through the moment-map difference and the metric normalizations at the two fixed points.
The ambiguity structure and infrared scaling of the bion weight are common to these targets, while the inferred pole position depends on the target metric.

\end{abstract}

\thispagestyle{empty}
\newpage\setcounter{page}{1}

\tableofcontents

\section{Introduction}\label{sec:intro}

The relation between divergent perturbation theory and nonperturbative effects is a central question in asymptotically free quantum field theories. Infrared renormalons give singularities on the positive real axis of the Borel plane, making the resummation of the perturbative series ambiguous \cite{tHooft:1977xjm,Beneke:1998ui}. Since physical observables should be independent of the prescription used to pass these singularities, the perturbative ambiguity must be accompanied by a corresponding ambiguity in the nonperturbative contribution \cite{Ecalle,JZJ}. Two-dimensional nonlinear sigma models provide a setting in which this cancellation and its relation to the infrared dynamics can be examined in a more concrete manner, see for instance, \cite{Fujii:2015bua,Basar:2015xna,Basar:2017hpr,Behtash:2015loa,Dunne:2015ywa,Cherman:2013yfa,Demulder:2016mja,Schepers:2020ehn,Marino:2021six,Marino:2021dzn,Marino:2022ykm,Liu:2025bqq,Marino:2024uco,DiPietro:2021yxb,Couso-Santamaria:2014iia,Fujimori:2018kqp,Hongo:2018rpy,Schubring:2021hrw,Dunne:2012ae}.

One approach is to use exact results at large $N$, where perturbative and nonperturbative contributions can be examined within the same calculation. Early studies of two-dimensional sigma models clarified the cancellation of infrared divergences \cite{David:1980rr} and the relation between infrared renormalons and nonperturbative effects in the $1/N$ expansion \cite{David:1982qv}. In the $O(N)$ sigma model, the operator product expansion provides a description in which ambiguities in the coefficient functions are compensated by those in the vacuum expectation values of composite operators \cite{David:1983gz,David:1985xj,Beneke:1998eq}. This cancellation was also demonstrated for the bosonic and fermionic two-point functions in the supersymmetric model at subleading order in $1/N$ \cite{Schubring:2021hrw}. Related results include the disappearance of infrared renormalon ambiguities in the full mass gap of the Gross-Neveu model at order $1/N$ \cite{Kneur:2001dd}. Moreover, calculations of the ground state energy in an external potential, both in the sigma model and its supersymmetric extension, allow the large-$N$ perturbative series to be compared with Bethe ansatz predictions and the associated renormalon contributions to be determined \cite{Marino:2021six}. A complementary question is whether the relevant nonperturbative ambiguity can be identified directly with saddle points in a semiclassical path integral.

Compactification on $\RR\times S^1$ with twisted boundary conditions makes such a semiclassical calculation possible in a weakly coupled regime \cite{Argyres:2012ka,Argyres:2012vv,Dunne:2012ae,Dunne:2015ywa}. In the $\cp^{N-1}$ models, the twist gives isolated vacua connected by fractional instantons, and a fractional instanton paired with an anti-fractional instanton forms a topologically neutral bion. The constituent separation and relative phase are quasi-moduli, and their integration can be defined using Lefschetz thimbles in the complexified configuration space. This prescription has been used to evaluate real and complex bion contributions and their imaginary ambiguities \cite{Fujimori:2016ljw,Fujimori:2017oab,Fujimori:2017osz}. In the two-dimensional theory, the KK fluctuations further renormalize the bion action and relate its exponential weight to the dynamically generated scale \cite{Fujimori:2018kqp}. The relation to perturbative renormalons nevertheless requires attention to the observable and the limit under consideration.

In this paper, we study this question first in the one-parameter deformation of the $\cp^1$ sigma model constructed from a Lie-algebraic metric \cite{Sheu:2023hoz,Gamayun:2023sif,Gamayun:2023atu,Sheu:2024sxf,Bykov:2020llx,Bykov:2020nal}. The deformation changes the target metric while preserving its topology and a $U(1)$ isometry. Its quantum-mechanical reduction has been examined near the supersymmetric point, including the contributions from real and complex bions \cite{Sheu:2024hae}. Here we consider the two-dimensional theory on $\RR\times S^1$ and introduce a source proportional to $\delta\epsilon$, correlating to the induced potential from fermion ground state in the quantum mechanic setup \cite{Sheu:2024hae}, to probe the vacuum energy around the $\cN=(2,2)$ supersymmetric theory. The calculation includes the bosonic and fermionic one-loop determinants and the integration over the relative quasi-moduli.

The ultraviolet terms reproduce the renormalization of both the constituent actions and their overlap interaction. For the Lie-algebraic family, this agreement requires the running of both the coupling and the elongation parameter. The fermionic determinant shifts the coefficient of the linear separation term, while the quasi-moduli integral has the same functional form as in the undeformed model. Its thimble decomposition gives a vanishing one-bion contribution at the supersymmetric point, a real correction at first order in $\delta\epsilon$, and the first imaginary ambiguity at second order. Taking $g_R^2$ as the coupling of the asymptotic expansion then determines the positive-axis perturbative pole required to cancel this ambiguity. The pole is inferred from this cancellation. A direct perturbative Borel calculation at finite radius is not performed here.

The calculation also admits a geometric generalization to a smooth compact K\"ahler target of one complex dimension with the topology of $S^2$ and a nontrivial $U(1)$ isometry. At leading dilute one-loop order, the target dependence of the final bion amplitude is encoded in the moment-map difference between the two fixed points and the product of their metric normalizations. The local $U(1)$ action fixes the linear separation term and the infrared scaling of the bion weight, while the ambiguity structure follows from the common quasi-moduli dependence. The numerical Borel-pole position in an expansion in $g_R^2$ nevertheless depends on the metric through the constituent action. This separates the geometric dependence of the semiclassical result from the features shared by the targets considered here.

The paper is organized as follows. Section~\ref{sec:rev} introduces the deformed $\cp^1$ model, its twisted compactification, and the real and complex bion solutions. Section~\ref{sec:bionfluc} evaluates the one-loop determinants and the quasi-moduli integral, and discusses the resulting ambiguity and inferred perturbative pole. The calculation is extended to a general one-complex-dimensional K\"ahler target with the stated $U(1)$ isometry in Section~\ref{sec:gen2K}.

\section{The deformed $\cp^1$ sigma model}\label{sec:rev}

In this section, we review the one-parameter deformation of the $\cp^1$ model constructed from the Lie-algebraic consideration \cite{Sheu:2023hoz,Gamayun:2023sif,Gamayun:2023atu}.

\subsection{Defining the model}

The deformed $\cp^1$ model\footnote{Here we construct the deformed model from the Lie-algebraic point of view. Conventionally, it is known to be the so-called sausage model \cite{Fateev:1992tk} and is associated to the $\eta$-deformation of the $\cp^1$ model \cite{Klimcik:2002zj, Klimcik:2008eq,Sfetsos:2015nya, Delduc:2013fga, Bykov:2020llx}. See also recent studies \cite{Bykov:2020llx,Affleck:2021vzo,Bonelli:2026flm} on this model and the references therein.} arises from the Lie-algebraic metric of $sl(2) \times sl(2)$ algebra with its non-linear realization, say,
\begin{align}
    T^+ = -\varphi^2\pdv{\varphi} \,,\quad 
    T^0 = \varphi\pdv{\varphi} \,,\quad
    T^- = \pdv{\varphi}
\end{align}
and their complex conjugates. $T^{\pm}$ and $T^0$ are the generators of $sl(2)$ algebra. Assuming the $U(1)$ symmetry of the system is preserved, we can construct a generic metric corresponding to a linear combination of quadratic operators
\begin{align}
    \cG^{1\bar{1}} \pdv{\varphi}\pdv{\bphi{}} 
    = n_1 T^-\overline{T}^- + n_2 T^0\overline{T}^0 + n_3 T^+\overline{T}^+ \,.
\end{align}
In addition, as long as the parameters $n_{1,3}$ are non-singular (i.e. neither 0 nor $\infty$), the metric $\cG_{1\bar{1}}$ can be simplified by rescaling $\varphi$ and $\bphi{}$. Namely,
\begin{align}\label{eq:dcpmetric}
    \cG_{1\bar{1}} = \frac{G}{g^2}\,,\qquad
    G := \frac{1}{1 + 2k \abs{\varphi}^2 + \abs{\varphi}^4}
\end{align}
i.e. 
\begin{align}
    n_1 = n_3 = g^2
    \qand
    n_2 = 2g^2k
\end{align}
with $k$ being the elongation constant. 
Here $G$ denotes the metric with the overall coupling factored out, so that the bosonic kinetic term is $\cL_B=(G/g^2)\d_\mu\vp\d_\mu\phib$.\footnote{This is the coupling convention of \cite{Fujimori:2018kqp}. The coupling used in \cite{Sheu:2024hae,Sheu:2024sxf} is related to ours by $g_{\rm QM}^2=2g^2$.}

\begin{figure}[t] 
  \centering 
  \includegraphics[width=.5\linewidth]{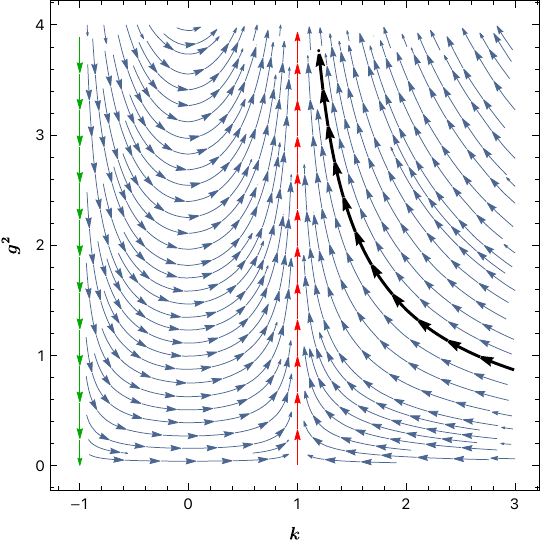}
  \caption{\small The RG flow of $g^2$ and $k$ couplings. The flow is from UV to IR. The separatrices are marked with green and red lines while the thick black line represents an example of the RG flow in the limit of the  sausage model  (i.e. large $k$). See for example \cite{Fateev:1992tk,Lukyanov:2019asr} for the illustration of the sausage model.}
  \label{fig:rgflow}
\end{figure}

Including the quantum corrections, the Lie algebraic sigma model is renormalizable \cite{Sheu:2023hoz,Gamayun:2023atu,Gamayun:2023sif} in two dimensions with the (one-loop) renormalization group (RG) flow equations in $g^2$ and $k$
\begin{align}\label{eq:RGE}
    \mu\dv{g^2}{\mu} = -\frac{kg^4}{\pi} \,,\quad
    \mu\dv{(g^2k)}{\mu} = -\frac{g^4}{\pi}\,.
\end{align}
Two coupling constants are entangled and the combination $g^4(k^2-1)$ is a RG-invariant.
Figure \ref{fig:rgflow} illustrates the direction of the renormalization group (RG) flow from the ultraviolet (UV) scale to the infrared (IR) scale. The separatrices are located at $k=\pm 1$, and in particular, the streamline at $k=1$ corresponds to the standard $\cp^1$ model.
In the $k \geq 1$ sector, the initial state with a small $g^2$ coupling follows the RG flow to a state with strong coupling. On the contrary, when the RG flow starts between the separatrices ($-1\leq k \leq 1$), the coupling initially weakens but eventually becomes strong as the elongation parameter approaches unity. The theory is asymptotically free and restore the full $O(3)$ symmetry in the IR regime.

\subsubsection*{Supersymmetric deformed $\cp^1$ model}

Since we will consider perturbations about the $\cN=(2,2)$ supersymmetric point, in analogy with the quantum-mechanical example \cite{Sheu:2024hae}, let us first describe the role of the fermions and the deformed potential. The fermionic sector of a general supersymmetric nonlinear sigma model can be written as
\begin{align}\label{eq:fermL}
\cL = \frac{2}{g^2}\left[
G\left( 
% \d\vp\db\phib + \d\phib\db\vp 
\frac{1}{2}\d_{\mu}\vp\d_{\mu}\phib 
- \psib_{L}D\psi_{L} - \psib_{R}\Db\psi_{R}
\right)
+ \frac{k+2\abs{\vp}^2+k\abs{\vp}^4}{(1+2k\abs{\vp}^2+\abs{\vp}^4)^3} \psi_{L}\psib_{L}\psi_{R}\psib_{R}
\right]
\end{align}
where
\begin{align}
D\psi_{L} := \left( \d+\Gamma\d\vp \right)\psi_{L} 
\,,\quad
\Db\psi_{R} := \left( \db+\Gamma\db\vp \right)\psi_{R} 
\,,\quad
\Gamma = -\frac{2\phib(k+\abs{\vp}^2)}{1+2k\abs{\vp}^2+\abs{\vp}^4}
\end{align}
The coefficient of the four-fermion interaction $\psi_{L}\psib_{L}\psi_{R}\psib_{R}$ is determined by the Riemann curvature $R_{1\bar{1}1\bar{1}}$.
Fermions do not contribute to the beta function at one loop. Their first contribution arises at two loops. The beta function is \cite{Sheu:2024sxf}
\begin{align}
    \beta(\cG) := \mu\dv{\cG}{\mu}
    = \frac{1}{4\pi}\cG\cR + \frac{2-N_f}{32\pi^2}\cG\cR^2 + \cdots
\end{align}
where $\cR$ stands for the scalar curvature of the full metric $\cG$ and $N_f$ is the number of Weyl fermions.\footnote{In the present convention,
\begin{align}
\cR = -2\cG^{-1}\partial_{\vp}\partial_{\phib}\log\cG
= 4g^2\frac{k+2\abs{\vp}^2+k\abs{\vp}^4}{1+2k\abs{\vp}^2+\abs{\vp}^4}\,.
\end{align}
The geometric coefficients are unchanged by the coupling redefinition. In particular, at $k=1$ the curvature is $\cR=4g^2$, and the corresponding coupling beta function reads
\begin{align*}
\beta(g^2) := \mu\dv{g^2}{\mu}
= -\frac{g^4}{\pi}-\frac{2-N_f}{2\pi^2}g^6+\cdots\,.
\end{align*}}
For $N_f=2$, which corresponds precisely to the $\cN=(2,2)$ theory, the two-loop coefficient vanishes as it is well known that the beta function is is exhausted at one loop when the interaction $R_{1\bar{1}1\bar{1}}(\bpsi{}\psi)^2$ is included.\footnote{At two loops, this vertex plays no role in determining the beta function, too. Examples of exact and perturbative beta functions for the $\cp^{N-1}$ model, its deformations, and homogeneous spaces can be found in \cite{Cui:2010si,Cui:2011uw,Chen:2014efa,Sheu:2019eis}.}

\subsection{On $\RR \times S^1$ and the deformed potential}

To study nonperturbative contributions to physical observables, we place the theory on the cylinder $\RR\times S^1$, where weak-coupling semiclassical calculations are tractable \cite{Argyres:2012ka,Argyres:2012vv,Dunne:2012ae}. We denote Euclidean time by $x$ and the coordinate along the circle of radius $R$ by $y$. The complex scalar and fermions obey the same twisted boundary condition,
\begin{align}
\vp(x,y+2\pi R) = e^{2\pi im R}\vp(x,y)
\,,\quad
\psi_{L,R}(x,y+2\pi R) = e^{2\pi im R}\psi_{L,R}(x,y)
\end{align}
where the dimensionless twist parameter $m$ satisfies $0<mR<1$. Equivalently, $m$ can be interpreted as a nontrivial background holonomy \cite{Dunne:2012ae,Eto:2004rz,Eto:2006mz,Eto:2006pg}. See also the recent review \cite{Poppitz:2021cxe}.\footnote{In the small-radius limit, the theory reduces to quantum mechanics, in which $m$ becomes the scalar-field mass; see, for example, \cite{Fujimori:2016ljw,Fujimori:2017oab,Fujimori:2017osz,Sheu:2024hae}.}
Keeping the lightest mode in the KK decomposition, $\vp(x,y)=\vp_0(x)e^{imy}$, yields the effective potential
\begin{align}
V = \frac{m^2}{g^2}\frac{\abs{\vp_0}^2}{1+2k\abs{\vp_0}^2+\abs{\vp_0}^4}
\end{align}
As in the undeformed $\cp^1$ model, this potential has two degenerate, isolated minima at $\vp_0=0$ and $\vp_0=\infty$, corresponding respectively to the north and south poles of the deformed sphere.

The twisted theory admits a BPS fractional instanton, which appears in the dimensionally reduced theory as a kink interpolating between the two minima. Its profile takes the same form as in the undeformed model \cite{Eto:2004rz,Eto:2006mz,Eto:2006pg,Bruckmann:2007zh,Brendel:2009mp,Bruckmann:2018rra},
\begin{align}
\vp_K(x,y) = a e^{m(x+iy)}
\,,\quad a = e^{-mx_0+i\alpha}
\end{align}
where $x_0$ and $\alpha$ are the position and phase moduli, respectively. Although the profile is independent of the elongation parameter $k$, the fractional-instanton action depends on it, say,
\begin{align}
S_k = \frac{2\pi mR}{g^2}
\cdot \frac{\arccosh{k}}{\sqrt{k^2-1}}
\end{align}
The factor $\arccosh{k}/\sqrt{k^2-1}$ varies smoothly with $k$ and approaches unity in the undeformed limit $k\to1$, in agreement with \cite{Sheu:2024sxf,Sheu:2024hae}. A bion configuration is then formed by a fractional instanton and an anti-fractional instanton, corresponding to a kink--antikink pair \cite{Fujimori:2018kqp}.

As our goal is to understand the bion contribution in the model, it is convenient to introduce a deformed potential to probe it as a perturbation around the supersymmetric point. That is, one can add the term 
\begin{align}\label{eq:defmp}
\delta{\cL} = -\frac{\delta{\epsilon} m}{2\pi R}\frac{1-\abs{\vp}^4}{1+2k\abs{\vp}^2+\abs{\vp}^4}
=: -\frac{\delta{\epsilon} }{2\pi R} \Delta{\mu}
\end{align}
Here $\Delta := G^{-1}\partial_{\vp}\partial_{\phib}$. 
This deformation can be derived from the moment map 
\begin{align}
\mu = \frac{m}{2\sqrt{k^2-1}}\log(\frac{\abs{\vp}^2+k-\sqrt{k^2-1}}{\abs{\vp}^2+k+\sqrt{k^2-1}})
+ \frac{m \log(k+\sqrt{k^2-1})}{\sqrt{k^2-1}}
\end{align}
with respect to the $U(1)$ symmetry of $\vp$. The model with a more general form of the moment map will be discussed in \cref{sec:gen2K}.
Note that in the reduced quantum mechanics this potential is naturally understood as the potential induced by fermions in the ground state sector \cite{Sheu:2024hae,Fujimori:2016ljw}.

We next outline the path-integral calculation of the vacuum-energy corrections induced by the source term $\Delta{\mu}$, following \cite{Fujimori:2018kqp}. Since \eqref{eq:defmp} is the only supersymmetry-breaking term, the vacuum energy can be expanded in powers of $\delta{\epsilon}$ as
\begin{align}
E(\delta{\epsilon}) = \sum_{j=0}^{\infty} E^{(j)}\left( \delta{\epsilon} \right)^{j}
= E^{(0)} + E^{(1)}\delta{\epsilon} + E^{(2)}(\delta{\epsilon})^{2} + \cdots
\end{align}
where $E^{(0)}=0$ by supersymmetry. The first two corrections will be needed below, since the leading bion ambiguity appears at second order in $\delta\epsilon$. To extract these coefficients, regulate the Euclidean-time volume by compactifying the $x$-direction as $x\sim x+\beta$. The partition function then takes the form
\begin{align}
Z(\delta{\epsilon})
&:= \int\cD{\phi_i} \;\exp\left[ -\int\dd[2]{x}\left( \cL + \delta{\cL} \right) \right]
\nonumber\\[2mm]
&= \int\cD{\phi_i} \;e^{-\int\cL\dd[2]{x}}
\Biggl[1+\frac{\delta{\epsilon}}{2\pi R}\int\dd[2]{x}\,\Delta{\mu}
+\frac{(\delta\epsilon)^2}{2(2\pi R)^2}
\left(\int\dd[2]{x}\,\Delta\mu\right)^2
+\order{\delta\epsilon^3}\Biggr]
\end{align}
where $\phi_i$ collectively denotes the scalar and fermionic fields. The vacuum energy follows from
\begin{align}
E(\delta{\epsilon})=-\lim_{\beta\to\infty}\frac{1}{\beta}\log Z(\delta{\epsilon})
\end{align}
Differentiating once with respect to $\delta\epsilon$ gives
\begin{align}\label{eq:1stE}
E^{(1)}
&= -\lim_{\beta\to\infty}\frac{1}{\beta}
\eval{\pdv{\delta{\epsilon}}\log{Z(\delta{\epsilon})}}_{\delta{\epsilon}=0}
\nonumber\\[2mm]
&= -\frac{1}{2\pi R}\lim_{\beta\to\infty}\frac{1}{\beta}
\int\dd[2]{x}\,\expval{\Delta{\mu}(x,y)}_0
= -\expval{\Delta{\mu}}_0
\,.
\end{align}
Here the subscript $0$ indicates that the expectation value is evaluated in the undeformed supersymmetric theory. Similarly, the second-order coefficient is
\begin{align}\label{eq:2ndE}
E^{(2)}
&=-\lim_{\beta\to\infty}\frac{1}{2\beta}
\eval{\pdv[2]{\delta\epsilon}\log Z(\delta\epsilon)}_{\delta\epsilon=0}
\nonumber\\[2mm]
&=-\frac{1}{4\pi R}\int_{\RR\times S^1}\dd[2]{x}\,
\left[\expval{\Delta\mu(x,y)\Delta\mu(0,0)}_0
-\expval{\Delta\mu}_0^2\right].
\end{align}
At weak coupling, the leading bion contributions to both $E^{(1)}$ and $E^{(2)}$ can be obtained by evaluating the fluctuations at one loop and integrating over the quasi-moduli. The latter integration produces an unambiguous first-order contribution and an imaginary ambiguity at second order, as discussed below. We therefore first identify the real and complex saddles that enter these corrections.

\subsection{Real and complex bion solutions on $\RR\times S^1$}

We now present the real and complex bion saddles that give the leading nonperturbative contribution to the path integral. The deformation $\delta{\cL}$ lifts the degeneracy of the two minima and selects $\vp=0$ as the lower-energy vacuum. The relevant bion configurations therefore approach $\vp=0$ as $x\to\pm\infty$ while making an excursion toward the second minimum at $\vp=\infty$. Following the construction of \cite{Sheu:2024hae}, the real bion solution is
\begin{align}\label{eq:realbion}
\varphi_{\rm rb}(x,y) = \sqrt{\frac{k\omega_k^2}{\omega_k^2-m^2}}\frac{e^{i(my+\alpha)}}{\sinh(\omega_k(x-x_0))}
\end{align}
where $x_0$ and $\alpha$ are the position and $U(1)$ phase moduli, respectively. The corresponding moduli space is therefore the cylinder $\RR\times S^1$. The divergence of \eqref{eq:realbion} at $x=x_0$ is a coordinate singularity. It simply indicates that the trajectory passes through the south pole $\vp=\infty$ of the target space. The frequency is
\begin{align}
    \omega_k = m \sqrt{1 + \frac{k\delta{\epsilon} g^2}{\pi mR}}
    \,.
\end{align}
To demonstrate its constituent structure, the real bion can be rewritten as a kink--antikink pair,
\begin{align}\label{eq:rbkak}
    \varphi_{\rm rb} = e^{imy}\left[
        e^{\omega_k(x - x_+) - i\alpha_+} + e^{-\omega_k(x - x_-) - i\alpha_-}
    \right]^{-1}
\end{align}
with $\bphi_{\rm rb}$ given by its complex conjugate, as in the undeformed $\cp^1$ model \cite{Fujimori:2016ljw,Fujimori:2017oab,Fujimori:2017osz,Fujimori:2018kqp}. Here $(x_\pm,\alpha_\pm)$ are the positions and phases of the kink and antikink, respectively, with
\begin{align}\label{eq:rbr}
    x_{\pm} = x_0 \pm \frac{1}{2\omega_k}\log\frac{4k\omega_k^2}{\omega_k^2-m^2}
    \,,\quad
    \alpha_{\pm} = \left( \alpha - \frac{\pi}{2} \right) \pm \frac{\pi}{2}
    \,.
\end{align}
Their separation is
\begin{align}
x_r := x_+-x_- = \frac{1}{\omega_k}\log\frac{4k\omega_k^2}{\omega_k^2-m^2}
\sim \frac{1}{m}\log\frac{1}{g^2}
\end{align}
at weak coupling. Thus, the bion becomes a well-separated kink--antikink pair in the semiclassical limit. This representation will also provide the bion ansatz for the subsequent valley analysis.

Evaluating the action on the real bion gives
\begin{align}\label{eq:rbaction}
    \srb = \frac{2\pi R\omega_k}{g^2\sqrt{k^2-1}}
    \left[ a_+^{-1}\log(\frac{a_++1}{a_+-1}) - a_-^{-1}\log(\frac{a_-+1}{a_--1})\right]
\end{align}
where
\begin{align}
    a_\pm = \frac{\sqrt{k} \omega_k}{ \sqrt{km^2 \pm\sqrt{k^2-1}(\omega_k^2-m^2)}}
    \,.
\end{align}
The contribution of this saddle to the vacuum energy is weighted by $e^{-\srb}$. In the weak-coupling limit, its action has the expansion
\begin{align}
    \srb = \srbz + \order{g^0}
    = \frac{4\pi mR}{g^2}\frac{\arccosh{k}}{\sqrt{k^2-1}}
    + \order{g^0}
    \,.
\end{align}
The leading term $\srbz$ is twice the fractional-instanton action, as expected for a kink--antikink configuration.

The semiclassical expansion must also include complex saddles. We complexify the field space by treating $\vp$ and $\bphi{}$ as independent fields, denoted by $\vp$ and $\tphi{}$. The resulting complex bion solution is \cite{Sheu:2024hae}
\begin{align}\label{eq:cpxbion}
    \varphi_{\rm cb}(x,y) = \sqrt{\frac{k\omega_k^2}{\omega_k^2-m^2}}
    \frac{e^{imy+i\alpha}}{i\cosh{\omega_k(x-x_0)}}
    % \nonumber\\[2mm]
    \,,\quad
    \tphi_{\rm cb}(x,y) = \sqrt{\frac{k\omega_k^2}{\omega_k^2-m^2}}
    \frac{e^{-imy-i\alpha}}{i\cosh{\omega_k(x-x_0)}}
    \,.
\end{align}
Unlike the real bion, this saddle does not satisfy the reality condition $\tphi{}=\bphi{}$. It nevertheless admits a kink--antikink representation,
\begin{align}\label{eq:cbkak}
\varphi_{\rm cb} &= e^{imy}\left[
    e^{\omega_k(x - \tilde{x}_+) - i\tilde{\alpha}_+} + e^{-\omega_k(x - \tilde{x}_-) - i\tilde{\alpha}_-}
    \right]^{-1}
\nonumber\\[2mm]
\tphi{}_{\rm cb} &= e^{-imy}\left[
    e^{\omega_k(x - \tilde{x}_+) + i\tilde{\alpha}_+} + e^{-\omega_k(x - \tilde{x}_-) + i\tilde{\alpha}_-}
    \right]^{-1}
\end{align}
where the complexified constituent parameters are
\begin{align}\label{eq:cbr}
    \tilde{x}_{\pm} = x_0 \pm \frac{1}{2\omega_k} \left( \log\frac{4k\omega_k^2}{\omega_k^2-m^2} -i\pi \right)
    \,,\quad
    \tilde{\alpha}_{\pm} = \alpha
    \,.
\end{align}
Relative to the real bion, the constituent separation is shifted by $-i\pi/\omega_k$. After analytic continuation of the coupling $g^2$, the complex-bion action has the same leading real part as \eqref{eq:rbaction} but acquires an imaginary part. The two choices of continuation give complex-conjugate contributions, whose ambiguity will enter the semiclassical evaluation of the vacuum energy.

\section{Nonperturbative contribution to vacuum energy}\label{sec:bionfluc}

This section evaluates the nonperturbative contributions of the real and complex bion saddles to the vacuum energy within the path-integral formalism. In the weak-coupling regime, the semiclassical expansion is retained through $\order{g^0}$, including the one-loop fluctuations about each saddle. The calculation requires determining the integration contour in the quasi-moduli space and evaluating the bosonic and fermionic one-loop determinants contributing to the vacuum energy.

\subsection{The quasi-moduli space}

The partition function, and hence the vacuum energy, can be evaluated using the Lefschetz-thimble prescription \cite{Witten:2010cx,Cristoforetti:2013wha,Fujii:2013sra,Tanizaki:2014tua,Tanizaki:2014xba,Kanazawa:2014qma,Tanizaki:2015pua,DiRenzo:2015foa,Fukushima:2015qza,Tanizaki:2015rda,Fujii:2015bua,Alexandru:2016gsd,Tanizaki:2016xcu}. This prescription provides a systematic construction of integration cycles in the complexified configuration space and relates their descriptions across different parameter regimes. In the weak-coupling limit $g^2\to0$, the massive fluctuations are integrated out at the quadratic order and encoded in one-loop determinants. The remaining functional integral then reduces to a finite-dimensional integral over the exact zero modes and quasi-zero modes.
More precisely, the bion has two exact zero modes: its center-of-mass position $x_0$ and overall $U(1)$ phase. The two quasi-moduli, $x_r$ and $\alpha_r$, describe the relative separation and relative phase of the constituent kink and antikink, respectively.

To evaluate the integral over the quasi-moduli space $\cM$, a family of bion configurations is first constructed using the valley method \cite{Balitsky:1986qn,Yung:1987zp,Aoyama:1991ca,Aoyama:1994sk,Aoyama:1995ca,Aoyama:1997qk,Aoyama:1998nt}. Let $\eta^a=(x_0,\alpha,x_r,\alpha_r)$ denote the collective coordinates on the real slice, which may equivalently be parametrized by the constituent positions and phases $(\xb{\pm},\ab{\pm})$. A valley configuration $\vp_B(\eta)$ is defined by requiring the functional gradient of the action to lie within the tangent space of $\cM$:
\begin{align}\label{eq:valleyeq}
\eval{\fdv{S}{\vp}}_{\vp=\vp_B}
= G\overline{K^a\pdv{\vp_B}{\eta^a}}
\end{align}
and its complex conjugate. Here $K^a$ is proportional to the action gradient along the collective-coordinate directions. The metric induced on $\cM$ is
\begin{align}
g_{a\bar b}
= \int\dd[2]{x}\;G
\pdv{\vp_B}{\eta^a}
\overline{\pdv{\vp_B}{\eta^b}}
\,.
\end{align}
Furthermore, the valley family passes through the real bion when $\eta^a$ takes the values in \cref{eq:rbr}. After analytic continuation of the collective coordinates, it also passes through the complex bion specified by \cref{eq:cbr}.
The fields are then decomposed into the bion background and their fluctuations,
\begin{align}\label{eq:expbf}
\mqty(\vp \\[2mm] \bar{\vp})
=
\mqty(\vp_B \\[2mm] \bar{\vp}_B)
+
\mqty( \delta{\vp} \\[2mm] \delta{\bar{\vp}} )
=
\mqty(\vp_B \\[2mm] \bar{\vp}_B)
+
gG^{-1/2}_B \mqty( \xi \\[2mm] \xib )
% \mqty( \xi_1 + i \xi_2 \\[2mm] \xi_1 - i \xi_2 )
\,,\quad
\psi_{L,R} = gG^{-1/2}_B \chi_{L,R}
\end{align}
where $G_B = G(\vp_B)$. Note that the fluctuation is orthogonal to the quasi-zero modes, say,
\begin{align}
\int G \bar{\pdv{\vp_B}{\eta^{\nu}}}\delta\varphi \dd[2]{x}
= \int G \pdv{\varphi_B}{\eta^{\nu}}\delta\bphi{}\dd[2]{x}
= 0 \,.
\end{align}
Using the decomposition \eqref{eq:expbf}, the action admits the semiclassical expansion
\begin{align}\label{eq:actexp}
S = \seff(\eta) -  \int \dd[2]{x} \Bigl\{ \cL^{(2)}(\xi,\xib) + 2\chib_L\nabla_{R}\chi_{L} + 2 \chib_{R}\nabla_{L}\chi_R  \Bigr\}
+ \order{g^2}
\end{align}
where $\seff(\eta)=S[\vp_B(\eta)]$ is the effective action of the bion field and $\cL^{(2)}(\xi,\xib)$ is the quadratic Lagrangian for the bosonic fluctuations. The term linear in the fluctuations vanishes because of the valley equation \eqref{eq:valleyeq} and the orthogonality condition. The leading quantum correction is therefore Gaussian and is determined entirely by the quadratic fluctuation operators. 

Introducing the fluctuation vector $(\xi,\xib)^T$, the quadratic bosonic action gives the linearized equation $\Delta_B\Xi=0$, where
\begin{align}
\Delta_{B} = \mqty(
\{\nabla_R,\nabla_L\}
& \cH/2
\\[2mm]
\bar{\cH}/2
& \{\nabla^{*}_{R},\nabla^{*}_{L}\}
)
+ \frac{1}{2}R_{1\bar{1}}\mqty(
\d_i\vp_B\d_i\phib_B
& -\d_{i}\vp_{B}\d_{i}\vp_B
\\[2mm]
-\d_{i}\phib_{B}\d_{i}\phib_B
& \d_i\phib_B\d_i\vp_B
)
+\order{g^2}
\end{align}
Here $R_{1\bar{1}}$ is the Ricci curvature of the targe manifold and the rest of notations are defined as
\begin{align}
&
\nabla_i := \d_i + 2iA_i \,,\quad
\nabla_R := \d+2iA_z 
\,,\quad 
\nabla_L := \db + 2iA_{\zb}
\,,\quad
\\[2mm]
&\nabla_i^{*} := \d_i - 2iA_i \,,\quad
\nabla_R^{*} := \d-2iA_z 
\,,\quad 
\nabla_L^{*} := \db - 2iA_{\zb}
\\[2mm]
&A_{i} = -\frac{i}{4}(\Gamma_B\d_i\vp_B - \bar{\Gamma}_B\d_i\phib_B) 
= \frac{i}{2} \cdot \frac{k+\abs{\vp_B}^2}{1+2k\abs{\vp_B}^2+\abs{\vp_B}^4} ( \phib_B\d_{i}\vp_B - \vp_B\d_i\phib_B)
\end{align}
and 
\begin{align}
\cH :=&~ 
\Gammab_{B}\left( \d_{i}+\Gamma_{B}\d_{i}\vp_{B} \right)\d_{i}\vp_{B}
\nonumber\\[2mm]
=&~ \frac{4\abs{\vp_{B}}^2(k+\abs{\vp_{B}}^2)^2}{(1+2k\abs{\vp_B}^2+\abs{\vp_{B}}^4)^2} \cdot \d_{i}\vp_{B}\d_{i}\vp_B
+ \frac{-2\vp_B(k+\abs{\vp_{B}}^2)}{1+2k\abs{\vp_B}^2+\abs{\vp_{B}}^4}\d_{i}\d_{i}\vp_{B}
\end{align}
In fact this differential operator is valid for all K\"ahler sigma models of complex dimension one with $U(1)$ symmetry.

Integrating out the bosonic and fermionic fluctuations at the quadratic order gives the one-bion contribution to the partition function,
\begin{align}
Z_1 = \int_{\cM}\dd[4]{\eta}\cJ\frac{\det\Delta_F}{\det'\Delta_B}e^{-\seff}
\end{align}
where $\cJ$ is the collective-coordinate Jacobian determined by the induced metric on $\cM$. The numerator contains the fermionic fluctuation determinant, while the prime on the bosonic determinant excludes the exact zero modes and quasi-zero modes already accounted for by the integration over $\eta$. Writing $Z=Z_0+Z_1+\cdots$, with $Z_0$ the contribution of the trivial-vacuum sector, 
\begin{align}\label{eq:piEt}
    E = \left( -\lim_{\beta \to \infty} \frac{1}{\beta} \log{Z_0} \right)
    - \lim_{\beta \to \infty} \frac{1}{\beta} \log( 1 + \sum_{p=1}^{\infty} \frac{Z_p}{Z_0})
\end{align}
where $Z_p$ is the $p$-bion saddle.
The leading bion contribution to the vacuum energy follows from expanding $\log Z$ to first order in $Z_1/Z_0$,
\begin{align}
E_1 = -\lim_{\beta\to\infty}\frac{1}{\beta}\frac{Z_1}{Z_0}
= -2\pi\int_{\cM_r}\dd{x_r}\dd{\alpha_r} \cJ\exp(-S_R)
\end{align}
Here $\cM_r$ denotes the space of relative quasi-moduli $(x_r,\alpha_r)$. The integrations over the center position and overall phase give factors of $\beta$ and $2\pi$, respectively, leaving the integral over $\cM_r$. Note that the divergence from the $x_0$ integral is regularized by temporarily compactifying the $x$ direction with the circumference $\beta$. The classical action and the fluctuation determinants, normalized to the trivial vacuum, are collected in
\begin{align}\label{eq:SR}
S_R(x_r,\alpha_r) = \seff(x_r,\alpha_r) - \log\frac{\det\Delta_F}{\det\Delta_F^0} + \log\frac{\det'\Delta_B}{\det\Delta_B^0}
\end{align}
where $\Delta_F^0$ and $\Delta_B^0$ are the fluctuation operators around the trivial vacuum.

\subsection{One-loop determinants}

The determinant calculation separates into the lowest KK sector, the nonzero bosonic KK modes, and the fermionic fluctuations. The following expressions describe the leading one-loop contribution, with $mx_r\gg1$ and $e^{-mx_r}$ of order $g^2$ at the relevant saddles. The logarithmically divergent overlap term is also needed to renormalize the interaction between the constituents.

\subsubsection*{Bosonic fluctuations}

The twisted boundary condition gives the Kaluza--Klein decomposition
\begin{align}\label{eq:detKK}
\xi(x,y)&=\sum_{n\in\ZZ}\xi_n(x)e^{i(n/R+m)y},
&
\xib(x,y)&=\sum_{n\in\ZZ}\tilde\xi_n(x)e^{i(n/R-m)y}.
\end{align}
The two-dimensional determinant therefore becomes a product of one-dimensional determinants. The $n=0$ sector contains the exact zero modes and quasi-zero modes already represented by the collective coordinates. Removing these modes, its determinant combines with the Jacobian to give \cite{Fujimori:2016ljw,Sheu:2024hae}
\begin{align}\label{eq:detBzero}
\cJ\left.\frac{\det\Delta_B^0}{\det'\Delta_B}\right|_{n=0}
=\left(\frac{4m^2R}{g^2}\right)^2
\end{align}
at leading order. Here $n=0$ labels a KK sector, rather than a vanishing eigenvalue and the remaining massive fluctuations in this sector are included in the determinant.

For $n\neq0$, the determinant ratios follow from the asymptotic solutions of the fluctuation equation \cite{Fujimori:2016ljw,Fujimori:2018kqp}. For a scalar channel $L_n=-\d_x^2+V_n(x)$ with $V_n(\pm\infty)=M_n^2$, let $f_n^-$ satisfy
\begin{align}
L_nf_n^-=0,
\qquad
\lim_{x\to-\infty}e^{-M_nx}f_n^-(x)=1.
\end{align}
The growing coefficient at the opposite end determines the normalized determinant,
\begin{align}\label{eq:detmatching}
\frac{\det L_n}{\det(-\d_x^2+M_n^2)}
=\lim_{x\to+\infty}e^{-M_nx}f_n^-(x).
\end{align}
A matrix version applies to coupled channels. In the $g^2\to 0$ limit, the leading equations separate into scalar channels near the constituents, so the required coefficients can be found by matching their solutions through the intermediate region.

Near the kink, the deformation enters through
\begin{align}\label{eq:WdCP}
W_-(x)=\frac{m\sinh[2m(x-x_-)]}{k+\cosh[2m(x-x_-)]}.
\end{align}
The factorized equation contains 
\begin{align}
\Delta_{B}\biggr|_{n} \sim \left( \d_x+\frac{n}{R}-W_- \right)\left( \d_x-\frac{n}{R}+W_- \right) 
\,,
\end{align}
whose two independent solutions can be written as
\begin{align}
f_{n,1}(x)&=\frac{e^{nx/R}}{\sqrt{k+\cosh[2m(x-x_-)]}},
\nonumber\\[2mm]
f_{n,2}(x)&=f_{n,1}(x)\int^x\dd{s}\,e^{-2ns/R}
\biggl\{k+\cosh[2m(s-x_-)]\biggr\}.
\end{align}
The antikink solutions follow by reversing the orientation. Matching the solutions and normalizing to the vacuum removes the center-position dependence. Although the individual profiles depend on $k$, the combined contribution of a pair of nonzero KK modes is
\begin{align}\label{eq:detBpair}
\left.\log\frac{\det\Delta_B}{\det\Delta_B^0}\right|_n
+\left.\log\frac{\det\Delta_B}{\det\Delta_B^0}\right|_{-n}
=2\log\frac{n-mR}{n+mR},
\qquad n>0,
\end{align}
at leading order in the separation expansion. This agrees with the undeformed $\cp^1$ result. The condition $0<mR<1$ ensures that the factors in this expression are positive.

The ultraviolet divergence is explicit in the KK product. With a cutoff $N$ on the positive KK index,
\begin{align}
\prod_{n=1}^{N}\frac{n-mR}{n+mR}
&=\frac{\Gamma(N+1-mR)}{\Gamma(N+1+mR)}
\frac{\Gamma(1+mR)}{\Gamma(1-mR)}
\sim N^{-2mR}\frac{\Gamma(1+mR)}{\Gamma(1-mR)}.
\end{align}
Identifying $N\simeq R\Lambda_0$, where $\Lambda_0$ is the ultraviolet cutoff, gives
\begin{align}\label{eq:detBnonzero}
\left.\log\frac{\det\Delta_B}{\det\Delta_B^0}\right|_{n\neq0}
=-4mR\log(R\Lambda_0)
+2\log\frac{\Gamma(1+mR)}{\Gamma(1-mR)}
\end{align}
up to overlap corrections. The first such correction is needed for the renormalization of the classical interaction. Its ultraviolet divergence can be extracted from the large-$n$ expansion of the fluctuation equation \cite{Fujimori:2018kqp}. After factoring out the vacuum exponential behavior, the remaining solution is expanded in powers of $1/n$. Only the $1/n$ term produces a logarithmic divergence upon summation, since 
\begin{align}
\sum_{n=1}^{N}\frac{1}{n}=\log N+\gamma_{E}+\order{N^{-1}}
\,,
\end{align}
where $\gamma_{E}$ is the Euler constant. Higher inverse powers give convergent sums. For a KK pair, this term is
\begin{align}
-\frac{R}{n}\int_{-\infty}^{\infty}\dd{x}\,
R_{1\bar1}\bigl(\abs{\d_x\vp_B}^2+m^2\abs{\vp_B}^2\bigr)
=-\frac{4mR}{n}\left(1-2k e^{-mx_r}\cos\alpha_r\right)
\end{align}
to first order in the overlap. Summing over $n$, the separation-independent term reproduces the divergence in \eqref{eq:detBnonzero}, while the overlap term gives
\begin{align}\label{eq:detBoverlap}
8kmR\log(R\Lambda_0)e^{-mx_r}\cos\alpha_r.
\end{align}
Its dependence on $e^{-mx_r}\cos\alpha_r$ matches the classical kink--antikink interaction, allowing this divergence to be absorbed into the coupling multiplying that interaction. In other words, this large-$n$ calculations coincide with the RG flow equation \eqref{eq:RGE},
\begin{align}\label{eq:detoverlaprenorm}
\frac{1}{g_R^2} = \frac{1}{g_{0}^{2}} - \frac{k}{\pi}\log\abs{R\Lambda_0} \,.
\end{align}

\subsubsection*{Fermionic fluctuations}

The fermionic determinant gives a separation-dependent contribution to the bion weight, which must be combined with the classical kink--antikink interaction. The background enters the fermion operators through the connection $A_i$. Its $x$-component can be removed from the one-dimensional equations by a phase redefinition,
\begin{align}
\chi_{L,R}(x,y)
=e^{-2i\int^x A_x(s)\dd{s}}
\sum_{n\in\ZZ}e^{i(n/R+m)y}\chi_{L,R,n}(x).
\end{align}
The remaining background dependence is therefore encoded in $A_y$, while the twist shifts the KK momentum to $n/R+m$. The normalized determinant of each KK sector is \cite{Fujimori:2018kqp}
\begin{align}
\left.\log\frac{\det\Delta_F}{\det\Delta_F^0}\right|_n
=2\,\sgn\left( \frac{n}{R}+m \right)\int_{-\infty}^{\infty}A_y(x)\dd{x}.
\end{align}
Only the sign of the shifted KK momentum enters this expression. Since $0<mR<1$, the sectors $n$ and $-n$ have opposite signs for every $n\geq1$, and their contributions cancel with a symmetric cutoff. The fermionic determinant thus reduces to the unpaired lowest KK sector $n=0$, namely,
\begin{align}\label{eq:detFconnection}
\log\frac{\det\Delta_F}{\det\Delta_F^0}
=2\int_{-\infty}^{\infty}A_y(x)\dd{x}.
\end{align}

To evaluate the connection integral, write $u_B=\vp_B\tphi_B$, with $\tphi_B=\bar\vp_B$ on the real slice. The constituent representation gives
\begin{align}
u_B&=\frac{e^{\omega_kx_r}}
{2[\cosh(2\omega_k(x-x_0))+\cos\alpha_r]},
&
A_y&=-m\frac{u_B(k+u_B)}{1+2ku_B+u_B^2}.
\end{align}
For a widely separated real pair, $u_B$ is large between the constituents and small outside them. Accordingly, $A_y$ approaches $-m$ in this region and vanishes far from the pair. The connection integral therefore contains a term proportional to the separation, 
\begin{align}
2\int A_y\dd{x}\simeq-2mx_r \,.
\end{align}
This also explains why its leading slope is independent of $k$: the limiting values of $A_y$ do not depend on the elongation parameter.
To determine the contributions from the constituent regions as well, introduce $q_\pm=k\pm\sqrt{k^2-1}$, for which $q_+q_-=1$. The connection then factorizes as
\begin{align}
A_y=-\frac{m}{2}
\left(\frac{1}{1+q_+/u_B}+\frac{1}{1+q_-/u_B}\right).
\end{align}
Its integral can be expressed through
\begin{align}
J(a,b)&:=\int_{-\infty}^{\infty}\frac{\dd{t}}{1+a(\cosh2t+b)}
=\frac{2}{\sqrt{(1+ab)^2-a^2}}
\arctanh\sqrt{\frac{1+a(b-1)}{1+a(b+1)}}.
\end{align}
This expression can be further analytically continued to complex quasi-moduli. As a result,
\begin{align}
\log\frac{\det\Delta_F}{\det\Delta_F^0}
=-\frac{m}{\omega_k}\sum_{\sigma=\pm}
J\left(2q_\sigma e^{-\omega_kx_r},\cos\alpha_r\right).
\end{align}
In the large-separation limit,
\begin{align}
J\left(2q_\pm e^{-\omega_kx_r},\cos\alpha_r\right)
=\omega_kx_r-\log q_\pm+\cdots.
\end{align}
The constant terms cancel since $q_+q_-=1$, and hence
\begin{align}\label{eq:detFfinal}
\log\frac{\det\Delta_F}{\det\Delta_F^0}=-2mx_r+\cdots.
\end{align}
Thus, the constant terms from the constituent regions cancel, leaving the leading fermionic determinant independent of $k$. 
On the real slice, the resulting factor $e^{-2mx_r}$ suppresses widely separated pairs and describes an attractive contribution to the kink--antikink interaction. Combining it with the classical deformation term $2m\delta\epsilon\,x_r$ gives 
\begin{align}
    2m\epsilon x_r 
    \qq{with}
    \epsilon=1+\delta\epsilon
    \,.
\end{align}

\subsubsection*{Renormalization and the combined factor}

So far, we have evaluated the one-loop determinants of the bosonic and fermionic fields around the bion background. Their ultraviolet divergences should reproduce the one-loop renormalization obtained in the two-dimensional calculation \cite{Sheu:2024sxf}, see \cref{eq:RGE}. To verify this agreement and assemble the finite bion weight, consider the classical valley action at the order needed here,
\begin{align}\label{eq:detclassical}
\seff
=\frac{4\pi mR}{g^2}
\left(\frac{\arccosh k}{\sqrt{k^2-1}}-2e^{-mx_r}\cos\alpha_r\right)
+2m\delta\epsilon\,x_r+\cdots.
\end{align}
The first term is the sum of the isolated constituent actions, while the term proportional to $e^{-mx_r}\cos\alpha_r$ describes their classical interaction. The ultraviolet contributions from the nonzero bosonic KK modes in \cref{eq:detBnonzero,eq:detBoverlap} add to the effective action as
\begin{align}\label{eq:1lprenom}
-4mR\log(R\Lambda_0)
\left(1-2k e^{-mx_r}\cos\alpha_r\right).
\end{align}
Thus, the separation-independent divergence renormalizes the constituent actions, and the overlap divergence renormalizes their interaction. The fermionic determinant contributes no additional ultraviolet divergence at this order, in agreement with the absence of a fermionic contribution to the one-loop beta function.

The overlap divergence is already accounted for by \eqref{eq:detoverlaprenorm}. To see how the separation-independent term in \eqref{eq:1lprenom} follows from the renormalization of both the coupling and the elongation parameter, differentiate the combination entering the constituent actions. Using \eqref{eq:RGE} gives
\begin{align}
\mu\dv{}{\mu}\left(\frac{1}{g^2}\frac{\arccosh k}{\sqrt{k^2-1}}\right)
&=-\frac{1}{g^4}\frac{\arccosh k}{\sqrt{k^2-1}}\mu\dv{g^2}{\mu}
+\frac{1}{g^2}\left[\frac{1}{k^2-1}
-\frac{k\arccosh k}{(k^2-1)^{3/2}}\right]\mu\dv{k}{\mu}
=\frac{1}{\pi}.
\end{align}
The terms proportional to $k\arccosh k/\sqrt{k^2-1}$ cancel between the two contributions, leaving a slope independent of $k$. Integrating from the cutoff $\Lambda_0$ to the scale $1/R$, with bare couplings $g^2,k$ and renormalized couplings $g_R^2,k_R$, respectively, yields
\begin{align}\label{eq:detsubtraction}
\frac{1}{g_R^2}\frac{\arccosh k_R}{\sqrt{k_R^2-1}}
&=\frac{1}{g^2}\frac{\arccosh k}{\sqrt{k^2-1}}
-\frac{1}{\pi}\log(R\Lambda_0).
\end{align}
Multiplying \cref{eq:detsubtraction} by the prefactor $4\pi mR$ therefore reproduces precisely the separation-independent correction $-4mR\log(R\Lambda_0)$ in \cref{eq:1lprenom}. Together with \cref{eq:detoverlaprenorm}, this absorbs both logarithmic terms into the classical action expressed in $g_R^2$ and $k_R$, demonstrating agreement with the two-dimensional one-loop renormalization.

After this subtraction, the nonzero bosonic KK modes leave the finite ratio of gamma functions in \eqref{eq:detBnonzero}. The fermionic determinant adds $2mx_r$ to the effective action and combines with the classical deformation through $\epsilon=1+\delta\epsilon$, as discussed above. Including the lowest-KK determinant and the quasi-moduli measure in \eqref{eq:detBzero}, the leading one-loop weight becomes
\begin{multline}\label{eq:detcombined}
\cJ e^{-S_R}
=\left(\frac{4m^2R}{g_R^2}\right)^2
\left[\frac{\Gamma(1-mR)}{\Gamma(1+mR)}\right]^2
\\[2mm]
\times\exp\left[
-\frac{4\pi mR}{g_R^2}\frac{\arccosh k_R}{\sqrt{k_R^2-1}}
+\frac{8\pi mR}{g_R^2}e^{-mx_r}\cos\alpha_r
-2m\epsilon x_r
\right].
\end{multline}
The prefactor collects the lowest-KK contribution and the finite nonzero-KK determinant, while the exponent contains the renormalized constituent actions, their overlap interaction, and the linear term induced by the deformation and fermions. Finite overlap corrections and higher-loop fluctuations lie beyond the leading semiclassical order considered here. This expression supplies the weight on the quasi-moduli space. The integration cycle and the associated imaginary ambiguity are determined by the Lefschetz-thimble analysis below.

\subsection{Lefschetz thimbles and determining IR renormalon poles}\label{sec:thimbles}

With all the ingredients obtained in the previous sections, we can now perform the integral over the quasi-moduli space via the Lefschetz-thimble prescription. The purpose is to determine which real and complex bion saddles contribute to the vacuum energy and how their contributions depend on the direction from which the real coupling is approached. Separating the relative quasi-moduli dependence in \eqref{eq:detcombined}, define
\begin{align}\label{eq:thimbleprefactor}
\cI=\left(\frac{4m^2R}{g_R^2}\right)^2
\left[\frac{\Gamma(1-mR)}{\Gamma(1+mR)}\right]^2
\exp(-\frac{4\pi mR}{g_R^2}\frac{\arccosh k_R}{\sqrt{k_R^2-1}}) \,.
\end{align}
The one-bion contribution to the vacuum energy is then
\begin{align}\label{eq:thimbleintegral}
E_1=-2\pi\cI\int_{\cM_r}\dd{x_r}\dd{\alpha_r}\,e^{-V_r},
\qquad
V_r=-\frac{8\pi mR}{g_R^2}e^{-mx_r}\cos\alpha_r+2m\epsilon x_r.
\end{align}
At this order, $k_R$ enters only through the separation-independent action in $\cI$. The quasi-moduli dependence is therefore identical to that in the undeformed model, and the contour analysis follows the same construction as in \cite{Fujimori:2016ljw,Fujimori:2018kqp,Sheu:2024hae}.

\subsubsection*{The quasi-moduli integral}

Since the quasi-moduli dependence in \eqref{eq:thimbleintegral} is the same as in the previous calculation \cite{Fujimori:2016ljw,Fujimori:2018kqp,Sheu:2024hae}, it suffices to only sketch the thimble construction here. The contour is determined by first introducing a small phase $g_R^2\to g_R^2e^{i\theta}$ and complexifying $x_r$ and $\alpha_r$. It is then deformed into a sum of middle-dimensional thimbles along which the integrand decays. With a suitable normalization of the flow parameter, the equations take the form
\begin{align}\label{eq:thimbleflow}
\dv{x_r}{t}=\frac{1}{2m\arccosh{k}}\overline{\pdv{V_r}{x_r}},
\qquad
\dv{\alpha_r}{t}=\frac{m}{2\arccosh{k}}\overline{\pdv{V_r}{\alpha_r}}.
\end{align}
As in \cite{Sheu:2024hae}, an overall positive factor in the flow metric can be absorbed by rescaling $t$. The imaginary part of $V_r$ remains constant along the flow, while its real part increases away from a saddle on the integration thimble.

For $\epsilon>0$, a branch flowing from the saddle towards large separation at fixed relative phase is
\begin{align}\label{eq:thimbletrajectory}
x_r(t)
&=\frac{1}{m}\log\left[
\frac{4\pi mR}{\epsilon g_R^2}
\left(1+e^{\epsilon m(t-t_0)}\right)\right]
+\frac{i}{m}(\sigma\pi-\theta),
\nonumber\\
\alpha_r(t)&=(1-\sigma)\pi,
\qquad \sigma\in\ZZ,
\end{align}
where $t_0$ fixes the origin of the flow parameter and the relative phase is understood modulo $2\pi$. The saddle is approached as $t\to-\infty$. At $\theta=0$, $\sigma=0$ corresponds to the real bion, while $\sigma=\pm1$ give the two complex continuations with imaginary separations $\pm\pi/m$. The separation involves $\epsilon$ because the fermionic determinant is already included in $V_r$.

The integration cycle is a linear combination $\cC_\theta=\sum_\sigma n_\sigma(\theta)\cJ_\sigma$, with the coefficients determined by the intersections of the original contour with the dual thimbles $\cK_\sigma$. These intersections are unchanged from the previous analysis \cite{Fujimori:2016ljw,Sheu:2024hae}: the cycle is $\cJ_0-\cJ_{-1}$ for $\theta\to0^-$ and $\cJ_1-\cJ_0$ for $\theta\to0^+$. Thus the contributing complex branch depends on the direction of the phase rotation, giving the Stokes ambiguity at real positive coupling.

The quasi-moduli integral can consequently be taken from the same calculation, with the interaction coefficient replaced by that in \eqref{eq:thimbleintegral}. The result is
\begin{align}\label{eq:qmthimbleresult}
\int_{\cC_\pm}\dd{x_r}\dd{\alpha_r}\,e^{-V_r}
=\frac{\pi}{m}\left(\frac{4\pi mR}{g_R^2}\right)^{-2\epsilon}
\frac{\Gamma(\epsilon)}{\Gamma(1-\epsilon)}e^{\mp i\pi\epsilon},
\end{align}
where the signs denote $\theta\to0^\pm$. The ratio of gamma functions and the phase include the combined contributions from the real and complex thimbles. In particular, their relative weights are essential for the cancellation at the supersymmetric point.

\subsubsection*{The one-bion contribution}

Together with the prefactor in \eqref{eq:thimbleprefactor}, this gives the leading one-bion contribution to the vacuum energy,
\begin{align}\label{eq:bionenergythimble}
E_{1,\pm}
&=-2m\left[\frac{4\pi mR}{g_R^2}\frac{\Gamma(1-mR)}{\Gamma(1+mR)}\right]^2
\exp(-\frac{4\pi mR}{g_R^2}
\frac{\arccosh k_R}{\sqrt{k_R^2-1}})
\cdot
\frac{\Gamma(\epsilon)}{\Gamma(1-\epsilon)}
\left(\frac{4\pi mR}{g_R^2}e^{\pm\frac{\pi}{2}i}\right)^{-2\epsilon} \,.
\end{align}
Note that the dependence on $k_R$ appears only in the constituent action, and the limit $k_R\to1$ reproduces the undeformed $\cp^1$ result \cite{Fujimori:2018kqp}. As in the determinant calculation, the result is valid at the leading one-loop order. The omitted corrections to the effective action and determinants contribute at higher semiclassical orders.

\subsubsection*{Imaginary ambiguity and the supersymmetric limit}

The two choices $\theta\to0^\pm$ give complex-conjugate results for real $g_R^2,k_R$, and $\epsilon$. Their imaginary parts can be read directly from \eqref{eq:bionenergythimble}. Namely, 
\begin{multline}\label{eq:bionimaginary}
\Im E_{1,\pm}
=\pm\frac{2m}{\pi}
\left[\frac{4\pi mR}{g_R^2}\frac{\Gamma(1-mR)}{\Gamma(1+mR)}\right]^2
\exp(-\frac{4\pi mR}{g_R^2}
\frac{\arccosh k_R}{\sqrt{k_R^2-1}})
\\[2mm]
\times 
\left(\frac{4\pi mR}{g_R^2}\right)^{-2\epsilon}
\Gamma(\epsilon)^2\sin^2(\pi\epsilon)
\,.
\end{multline}
The sign follows the direction of the phase rotation, with the upper sign corresponding to $\theta\to0^+$. This dependence is an ambiguity in the one-bion contribution. For the full vacuum energy to be real and independent of this choice, the perturbative sector must provide the opposite imaginary ambiguity, as expected from the resurgent description.

At the supersymmetric point $\epsilon=1$, the real and complex thimble contributions cancel, giving $E_1=0$ as required by supersymmetry. Each individual thimble integral is nonzero. 
The cancellation follows only after combining them along the prescribed cycle. To examine how the ambiguity appears away from this point, expand \eqref{eq:bionenergythimble} in $\delta\epsilon$,\footnote{One should not confuse the notations $E_1$ and $E^{(1)}$. The former denotes the one-bion contribution, whereas the latter denotes the first-order correction away from the supersymmetric point.}
\begin{multline}\label{eq:bionnearsusy}
E_{1,\pm}
=-2m\left[\frac{\Gamma(1-mR)}{\Gamma(1+mR)}\right]^2
\exp(-\frac{4\pi mR}{g_R^2}
\frac{\arccosh k_R}{\sqrt{k_R^2-1}})
\\[2mm]
\times\biggl\{\delta\epsilon
-2\left[\gamma_E+\log\frac{4\pi mR}{g_R^2}
\pm\frac{i\pi}{2}\right](\delta\epsilon)^2
+\order{(\delta\epsilon)^3}\biggr\}.
\end{multline}
The term linear in $\delta\epsilon$ is real and gives an unambiguous one-bion contribution to $E^{(1)}$. The first imaginary ambiguity arises at second order. Denoting the one-bion part of $E^{(2)}$ by $E_1^{(2)}$, one obtains
\begin{align}\label{eq:bionsecondambiguity}
\Im E_{1,\pm}^{(2)}
=\pm2\pi m\left[\frac{\Gamma(1-mR)}{\Gamma(1+mR)}\right]^2
\exp(-\frac{4\pi mR}{g_R^2}
\frac{\arccosh k_R}{\sqrt{k_R^2-1}})
\,.
\end{align}
Thus the ambiguity is absent at first order in the supersymmetry-breaking deformation and appears at order $(\delta\epsilon)^2$, in agreement with the nearly supersymmetric calculation in quantum mechanics \cite{Sheu:2024hae}.

\subsubsection*{The infrared scale and the associated Borel singularity}

To relate this ambiguity to the IR renormalon, consider first the scale dependence of the exponential in \eqref{eq:bionimaginary}. The RG relation derived in the preceding subsection allows the one-loop invariant scale to be written as
\begin{align}\label{eq:thimbleRGscale}
\Lambda:=\mu\exp(-\frac{\pi}{g^2(\mu)}
\frac{\arccosh k(\mu)}{\sqrt{k(\mu)^2-1}}).
\end{align}
Evaluating this relation at $\mu=1/R$ gives
\begin{align}\label{eq:bionIRscale}
\exp(-\frac{4\pi mR}{g_R^2}
\frac{\arccosh k_R}{\sqrt{k_R^2-1}})
=(\Lambda R)^{4mR}.
\end{align}
The exponent $4mR$ is therefore fixed by the one-loop RG flow and is independent of the elongation parameter. The dependence on the RG trajectory is contained in $\Lambda$.

To determine the corresponding location in the Borel plane, consider the perturbative expansion of $E^{(2)}$ in $g_R^2$ at fixed $k_R$, and write its Borel transform and the two lateral sums as
\begin{align}\label{eq:borelconvention}
E_{\rm pert}^{(2)}(g_R^2)
&\sim c_0+\sum_{n=0}^{\infty}c_{n+1}(g_R^2)^{n+1},
\qquad
\cB E_{\rm pert}^{(2)}(t):=\sum_{n=0}^{\infty}\frac{c_{n+1}}{n!}t^n,
\nonumber\\
\mathcal{S}_\pm E_{\rm pert}^{(2)}
&:= c_0+\int_0^{\infty e^{\pm i0}}\dd{t}\,
e^{-t/g_R^2}\cB E_{\rm pert}^{(2)}(t).
\end{align}
Here the coefficients depend on $k_R$ and $mR$, and the contours pass above or below the positive real Borel axis. Comparing the exponential in the Borel integral with \eqref{eq:bionsecondambiguity} identifies the singularity required to cancel the bion ambiguity at
\begin{align}\label{eq:bionsingulant}
t=t_{\rm b}:= 4\pi mR\frac{\arccosh k_R}{\sqrt{k_R^2-1}}.
\end{align}
At order $(\delta\epsilon)^2$, the leading ambiguity in \eqref{eq:bionsecondambiguity} contains no additional power of $g_R^2$ multiplying the exponential. In the convention of \eqref{eq:borelconvention}, this is the dependence produced by a simple pole. Matching its residue to the imaginary part of the one-bion contribution requires
\begin{align}\label{eq:predictedborelpole}
\cB E_{\rm pert}^{(2)}(t)
\sim\frac{2m}{t-t_{\rm b}}
\left[\frac{\Gamma(1-mR)}{\Gamma(1+mR)}\right]^2.
\end{align}
The contours passing above and below this pole are anticipated to give imaginary parts opposite to \eqref{eq:bionsecondambiguity}, so that the sum of the perturbative and one-bion contributions is unambiguous at this order. Note that \eqref{eq:predictedborelpole} is inferred from the required cancellation. Higher-order fluctuations can modify the singular behavior beyond the leading pole. For general $\epsilon$, the additional coupling power in \eqref{eq:bionimaginary} also permits a branch point, and the simple-pole form need not apply away from the nearly supersymmetric expansion.
In the zero radius limit, the cancellation was shown explicitly in the quantum mechanical case \cite{Sheu:2024hae}. 

In the undeformed limit $k_R\to1$, \eqref{eq:bionsingulant} gives $t_{\rm b}=4\pi mR$. For general $k_R$, the location changes through the geometric factor in \eqref{eq:bionsingulant}, even though the power of $\Lambda R$ remains the same.

Configurations containing $p$ neutral bions consist of $p$ fractional instantons and $p$ anti-fractional instantons, grouped into pairs of zero net topological charge. In the dilute limit, their leading actions add, giving the exponential weight $\exp(-p\,t_{\rm b}/g_R^2)$. For $p\geq2$, these weights indicate possible further Borel singularities at $p\,t_{\rm b}$. Establishing such singularities and their coefficients requires a multi-bion calculation including the interactions between the pairs. 
At the one-bion level considered here, the thimble integral determines the leading ambiguity on $\RR\times S^1$ and the perturbative singularity required to cancel it.

\section{General K\"ahler target spaces with $U(1)$ isometry}\label{sec:gen2K}

In this section, we generalize the preceding calculation to a nonlinear sigma model with a one-complex-dimensional K\"ahler target space admitting a $U(1)$ isometry. The explicit deformed $\cp^1$ metric entered the bion calculation through its moment map, the behavior near the two fixed points, and the fluctuation operators along the constituents. Expressing these ingredients geometrically makes it possible to carry over the calculation without repeating the valley and determinant analysis.

\subsection{Geometric setup and fractional instantons}

Consider a smooth compact K\"ahler target manifold $M$ of complex dimension one, topologically $S^2$, equipped with a $U(1)$ isometry and hence possessing two fixed points.
The generalization concerns its metric, which need not belong to the Lie-algebraic family. In a coordinate adapted to the $U(1)$ action, $\vp\to e^{i\theta}\vp$, the metric and K\"ahler form are
\begin{align}\label{eq:generalmetric}
\dd{s}^2=2G(u)\dd{\vp}\dd{\phib},
\qquad \Omega=iG(u)\dd{\vp}\wedge\dd{\phib},
\qquad u=\abs{\vp}^2.
\end{align}
As in \cref{eq:dcpmetric}, $G$ has the overall coupling factored out, and $\cG=G/g^2$ is the full kinetic metric. The Killing vector generating the twist has $X^\vp=im\vp$. 
With the moment map defined as 
\begin{align}
\dd{\mu}=-\iota_X\Omega
\end{align}
one has 
\begin{align}\label{eq:generalmm}
\partial_{\phib}\mu=-iGX^\vp=mG\vp
\implies
\mu(u) = m\int_{0}^{u}G(u')\dd{u'}
% \,,\qquad \dv{\mu}{u}=mG(u).
\,.
\end{align}
The twisted compactification therefore induces the potential
\begin{align}\label{eq:generalpotential}
V=\frac{1}{g^2}GX^\vp\overline{X^\vp}
=\frac{1}{g^2}G^{-1}\partial_{\vp}\mu\partial_{\phib}\mu
=\frac{m^2}{g^2}uG(u)
\,.
\end{align}
Since the metric is positive, $V\geq0$ and its minimum $V=0$ is attained precisely where the Killing vector vanishes. These points are fixed under the $U(1)$ action. The component $X^\vp=im\vp$ vanishes at $\{p_-:\vp=0\}$. To identify the other fixed point, we move to another chart $w=1/\vp$ near $\vp=\infty$. The Killing vector then becomes $X^w=(\partial w/\partial\vp)X^\vp=-imw$, which vanishes at $\{p_+:w=0\}$. Smoothness of the metric in this chart ensures that the potential also vanishes there. Thus $p_-$ and $p_+$ are the two degenerate classical vacua of the twisted theory, before the supersymmetry-breaking deformation is added.

The fractional instanton follows the gradient of the moment map,
\begin{align}\label{eq:generalkink}
\partial_x\vp=G^{-1}\partial_{\phib}\mu=m\vp,
\qquad
\vp_K(x,y)=e^{m(x-x_0)+i(my+\alpha)}.
\end{align}
Thus the fractional-instanton profile is unchanged, while its action depends on the metric through the difference of the moment map between the two vacua. The action of a fractional instanton takes the form
\begin{align}\label{eq:generalkinkaction}
S_K
&=\frac{2\pi R}{g^2}\int_{-\infty}^{\infty}\dd{x}\,
G\left(\abs{\partial_x\vp}^2+m^2\abs{\vp}^2\right)
\nonumber\\[2mm]
&=\frac{2\pi R}{g^2}\int_{-\infty}^{\infty}\dd{x}\,
G\abs{\partial_x\vp-m\vp}^2
+\left[ \frac{2\pi R}{g^2}\mu \right]_{x=-\infty}^{x=+\infty}
= \frac{2\pi R}{g^2}\Delta_{\mu} \,,
\end{align}
where the square term vanishes from the BPS eom. Here we denote the difference of the moment map at two vacua by 
\begin{align}
\Delta_\mu:= \mu(p_+)-\mu(p_-)
\end{align}
The anti-fractional instanton has the same action with the endpoints interchanged and the opposite BPS equation. The action of the kink-antiknik pari, i.e. a bion, is therefore $S_b=2S_K$.
For the deformed $\cp^1$ metric, $\Delta_\mu=m\arccosh k/\sqrt{k^2-1}$, reproducing the actions in \cref{sec:rev}. 

The supersymmetry-breaking deformation also has a geometric form. Using the same operator $\Delta=G^{-1}\partial_{\vp}\partial_{\phib}$ as in \eqref{eq:defmp} gives
\begin{align}\label{eq:generaldeformation}
\delta\cL=-\frac{\delta\epsilon}{2\pi R}\Delta\mu,
\qquad
\Delta\mu=-i\nabla_\vp X^\vp
=m\left(1+u\dv{\log G}{u}\right)
\,,
\end{align}
where $\nabla_{\vp}=\d_{\vp}+\Gamma$.
The endpoint values of $\Delta\mu$ follow from the regularity of the metric in the two charts. Near $p_-$, $\vp$ is regular, so smoothness and positivity give
\begin{align}\label{eq:generalnorthmetric}
G(u)=G_-+\order{u},
\qquad G_-=G(0)>0,
\qquad u\to0.
\end{align}
Near $p_+$, regularity is imposed in the $w=1/\vp$ chart, where the metric component is
\begin{align}\label{eq:generalsouthmetric}
G^{(w)}(w,\bar w)
&=G(u)\abs{\pdv{\vp}{w}}^2
=\abs{w}^{-4}G\left(\abs{w}^{-2}\right)
=G_++\order{\abs{w}^2},
\qquad G_+>0.
\end{align}
Since $u=\abs{w}^{-2}$, this implies
\begin{align}
G(u)=\frac{G_+}{u^2}\left[1+\order{u^{-1}}\right],
\qquad u\to\infty.
\end{align}
Thus the vanishing of $G(u)$ at large $u$ is due to the coordinate transformation and the metric remains nondegenerate at $p_+$. Substituting the two expansions into \eqref{eq:generaldeformation} gives
\begin{align}\label{eq:generalendpointsource}
(\Delta\mu)(p_-)
&=m\left[1+\lim_{u\to0}u\dv{\log G}{u}\right]=m,
\nonumber\\[2mm]
(\Delta\mu)(p_+)
&=m\left[1+\lim_{u\to\infty}u\dv{\log G}{u}\right]
=m(1-2)=-m.
\end{align}
The same result can be understood directly from the linear action of the isometry near each fixed point,
\begin{align}
X^\vp=im\vp \quad\text{near }p_-\,,
\qquad
X^w=\pdv{w}{\vp}X^\vp=-imw \quad\text{near }p_+\,.
\end{align}
At a fixed point, the connection term in $\nabla_\vp X^\vp$ vanishes because $X$ itself vanishes in the regular chart. The identity $\Delta\mu=-i\nabla_\vp X^\vp$ therefore gives the coefficients $+m$ and $-m$ of these linearized rotations. These are the two $U(1)$ weights, including the twist scale $m$, and are independent of the metric normalizations $G_\pm$ and the profile between the poles.

\subsection{Bion effective action and one-loop structure}

The endpoint coefficients $G_\pm$ in \cref{eq:generalnorthmetric,eq:generalsouthmetric} specify the normalization of the constituent tails. It is convenient to define an invariant combination
\begin{align}\label{eq:generalendpoints}
G_*:= \sqrt{G_-G_+}.
\end{align}
The two endpoint coefficients need not be equal. Their product fixes the lowest-KK prefactor, while their ratio enters the finite contributions from the constituent regions. Both dependences must be included when the metric is written without a rescaling of $\vp$.
The dilute pair has the same collective coordinates $(x_0,\alpha,x_r,\alpha_r)$ as before. Its leading action, measured relative to the vacuum at $p_-$, is
\begin{align}\label{eq:generalvalley}
\seff=\frac{4\pi R}{g^2}
\left(\Delta_\mu-2mG_+e^{-mx_r}\cos\alpha_r\right)
+2m x_r\delta{\epsilon} -\delta\epsilon\log\frac{G_+}{G_-}+\cdots.
\end{align}
The first term is twice the fractional-instanton action, while the overlap coefficient is fixed by $G_+$ through the tails near $p_+$, the vacuum approached between the constituents. The linear term follows directly from \eqref{eq:generaldeformation}. Between the constituents, $\Delta\mu$ differs from its vacuum value by $-2m$ over an interval of length $x_r$. The constituent regions also contribute a finite boundary term, so the complete leading source integral is
\begin{align}\label{eq:generalsourceintegral}
\int_{-\infty}^{\infty}\dd{x}\left(\Delta\mu-m\right)
=-2mx_r+\log\frac{G_+}{G_-}+\cdots.
\end{align}
Multiplying by $-\delta\epsilon$ gives the last two terms in \eqref{eq:generalvalley}. The combinations $G_+e^{-mx_r}$ and $-2mx_r+\log(G_+/G_-)$ are invariant under a rescaling of $\vp$ when the constituent positions are transformed accordingly.

The leading fermionic determinant is fixed by the same endpoint values. As in \cref{sec:bionfluc}, the paired nonzero KK sectors cancel and the remaining connection integral gives
\begin{align}\label{eq:generalfermion}
\log\frac{\det\Delta_F}{\det\Delta_F^0}
=-2mx_r+\log\frac{G_+}{G_-}+\cdots.
\end{align}
The same boundary term appears because the connection integral reduces to \eqref{eq:generalsourceintegral}. Subtracting this logarithm from the classical action shifts the coefficient of both the linear term and the endpoint constant from $\delta\epsilon$ to $\epsilon$.

For the bosonic determinant, the calculation near each fractional instanton admits the same factorization as in the explicit model. Up to an overall normalization that cancels in determinant ratios, the kink fluctuation operator takes the form
\begin{align}\label{eq:generalfactorization}
\Delta_{B,n}\sim
\left(\partial_x+\frac{n}{R}-W\right)
\left(\partial_x-\frac{n}{R}+W\right),
\qquad W=-\left.\Delta\mu\right|_{\vp_K}.
\end{align}
The antikink gives the opposite ordering of the two factors. The curvature term restores this factorization through $\d_xW=2R_{1\bar{1}}X^\vp\overline{X^\vp}$. For the deformed $\cp^1$ kink, it reproduces \cref{eq:WdCP}.

The asymptotic matching depends on the Killing-vector norm near the fixed points. The metric normalization factors cancel when the $n$ and $-n$ sectors are paired, leaving the same separation-independent contribution as in \cref{eq:detBnonzero},
\begin{align}\label{eq:generalboson}
\left.\log\frac{\det\Delta_B}{\det\Delta_B^0}\right|_{n\ne0}
=-4mR\log(R\Lambda_0)
+2\log\frac{\Gamma(1+mR)}{\Gamma(1-mR)}+\cdots.
\end{align}
The $\cdots$ stand for the contricutions beyond the leading semiclassical order. 
For the lowest KK modes and the Jacobian, the two endpoint amplitudes of each constituent give an extra factor $\sqrt{G_-G_+}$, so that
\begin{align}\label{eq:generalzeroprefactor}
\cJ\left.\frac{\det\Delta_B^0}{\det'\Delta_B}\right|_{n=0}
=\left(\frac{4m^2R}{g_R^2}\right)^2G_-G_+
=\left(\frac{4m^2RG_*}{g_R^2}\right)^2
\end{align}
at the leading order.

After absorbing the divergences into the running metric, the quasi-moduli potential, including the finite endpoint term, is therefore
\begin{align}\label{eq:generalrelative}
V_r=-\frac{8\pi mRG_+}{g_R^2}e^{-mx_r}\cos\alpha_r
+2m\epsilon x_r-\epsilon\log\frac{G_+}{G_-}.
\end{align}
Here and below, $G_\pm$ and $G_*$ in renormalized expressions are evaluated at the scale $1/R$. The endpoint constant multiplies the bion weight by $(G_+/G_-)^\epsilon$ and does not affect the flow equations. The dependence on $(x_r,\alpha_r)$ is unchanged, so the thimble calculation in Section~\ref{sec:thimbles} applies with the overlap coefficient proportional to $G_+$.

\subsection{Bion ambiguity and the renormalon}

The quasi-moduli integral in \eqref{eq:qmthimbleresult} now gives a power of $4\pi mRG_+/g_R^2$. Combining it with the endpoint constant in \eqref{eq:generalrelative} yields
\begin{align}\label{eq:generalendpointcancel}
\left(\frac{G_+}{G_-}\right)^\epsilon
\left(\frac{4\pi mRG_+}{g_R^2}\right)^{-2\epsilon}
=\left(\frac{4\pi mR\sqrt{G_-G_+}}{g_R^2}\right)^{-2\epsilon}
=\left(\frac{4\pi mRG_*}{g_R^2}\right)^{-2\epsilon}.
\end{align}
Together with the constituent action and \cref{eq:generalzeroprefactor}, this gives the expression of the one-loop correction to the vacuum energy
\begin{multline}\label{eq:generalenergy}
E_{1,\pm}
=-2m\left[\frac{4\pi mRG_*}{g_R^2}\frac{\Gamma(1-mR)}{\Gamma(1+mR)}\right]^2
\exp(-\frac{4\pi R}{g_R^2}\Delta_{\mu,R})
\cdot \frac{\Gamma(\epsilon)}{\Gamma(1-\epsilon)}
\left(\frac{4\pi mRG_*}{g_R^2}e^{\pm\frac{\pi}{2}i}\right)^{-2\epsilon}
\,,
\end{multline}
where $\Delta_{\mu,R}$ is the endpoint difference for the reference metric at $1/R$. At this leading dilute one-loop order, the target dependence is encoded in $\Delta_{\mu,R}$ and $G_*$. 

Expanding about the supersymmetric point gives
\begin{align}\label{eq:generalnearsusy}
E_{1,\pm}
=&-2m\left[\frac{\Gamma(1-mR)}{\Gamma(1+mR)}\right]^2
\exp\left(-\frac{4\pi R}{g_R^2}\Delta_{\mu,R}\right)
\nonumber\\[2mm]
&\times\left\{\delta\epsilon
-2\left[\gamma_E+\log\frac{4\pi mRG_*}{g_R^2}
\pm\frac{i\pi}{2}\right](\delta\epsilon)^2
+\order{(\delta\epsilon)^3}\right\}.
\end{align}
The one-bion contribution to $E^{(1)}$ is again real, while the first imaginary ambiguity occurs in $E^{(2)}$. The two lateral continuations are complex conjugates, and the cancellation at $\epsilon=1$ follows from the same real and complex thimble combination as before. Thus both the Stokes structure and the order in $\delta\epsilon$ at which the ambiguity appears extend to this class of targets.

To discuss the infrared scale, consider the one-loop RG flow of the metric $G/g^2$ appearing in the kinetic term. Let $\nu$ denote the renormalization scale, to distinguish it from the moment map. In the normalization of \cref{sec:rev},
\begin{align}\label{eq:generalRicci}
\pdv{}{\log\nu}\left(\frac{G}{g^2}\right)=\frac{1}{2\pi}R_{1\bar{1}},
\qquad
R_{1\bar{1}}=-\dv{}{u}\left(u\dv{\log G}{u}\right).
\end{align}
The running of the moment map follows from \eqref{eq:generalmm}. Namely,
\begin{align}\label{eq:generalmomentRG}
\pdv{}{\log\nu}\left(\frac{\mu}{g^2}\right)
=-\frac{1}{2\pi}\Delta\mu+c(\nu).
\end{align}
The additive term $c(\nu)$ is independent of the target coordinate and cancels in the difference between the two fixed points. Applying this relation to the constituent action gives
\begin{align}\label{eq:generalRGslope}
\dv{S_b}{\log\nu}
&=4\pi R\dv{}{\log\nu}
\left[\frac{\mu(p_+)}{g^2}-\frac{\mu(p_-)}{g^2}\right]
=-2R\left[(\Delta\mu)(p_+)-(\Delta\mu)(p_-)\right]
=4mR.
\end{align}
Here the endpoint values in \eqref{eq:generalendpointsource} have been used, with $m$ and $R$ held fixed. The detailed metric dependence thus drops out of the RG flow, leaving only the difference of the two $U(1)$ weights. Integrating from the cutoff $\Lambda_0$ to the scale $1/R$ gives
\begin{align}\label{eq:generalRGsubtraction}
S_{b,R}=S_{b,{\rm bare}}-4mR\log(R\Lambda_0),
\qquad e^{-S_{b,R}}=(\Lambda R)^{4mR},
\end{align}
where $\Lambda$ is the integration constant specifying the infrared scale along the RG trajectory. The first relation reproduces precisely the logarithmic correction in \cref{eq:generalboson}, while the second gives the same infrared scaling of the neutral-bion ambiguity as in \cref{eq:bionIRscale}.

Taking $g_R^2$ as the coupling of the asymptotic expansion at fixed reference metric, the exponential in \eqref{eq:generalnearsusy} determines the corresponding Borel singularity. In the convention of \eqref{eq:borelconvention}, cancellation of the one-bion ambiguity requires a perturbative singularity on the positive real axis at
\begin{align}\label{eq:generalsingulant}
t_{\rm b}=4\pi R\Delta_{\mu,R},
\qquad e^{-S_b}=e^{-t_{\rm b}/g_R^2}.
\end{align}
At order $(\delta\epsilon)^2$, the imaginary part has no additional coupling power, so the leading singular term is a simple pole, say \cref{eq:predictedborelpole}. Its residue gives the imaginary part opposite to that of the one-bion contribution in \eqref{eq:generalnearsusy}. The pole position depends on the metric through $\Delta_{\mu,R}$ and reduces to \eqref{eq:bionsingulant} for the deformed $\cp^1$ family, while the bion weight scales as $(\Lambda R)^{4mR}$ for all targets considered here. This leading pole is inferred from ambiguity cancellation; its direct verification requires the perturbative Borel transform.

\section{Discussions and conclusions}
\label{sec:clsn}

In this paper, we have studied the leading nonperturbative contributions to the vacuum energy of the deformed $\cp^1$ sigma model on $\RR\times S^1$, and extended the calculation to a smooth compact K\"ahler target of one complex dimension with a nontrivial $U(1)$ isometry and the topology of $S^2$. The twisted compactification gives two isolated vacua and admits fractional instantons connecting them. Introducing the supersymmetry-breaking source allows the vacuum energy to be expanded in $\delta\epsilon$, so that the contributions from real and complex bions can be calculated near the supersymmetric point in the weak-coupling limit.

For the deformed $\cp^1$ model, the bosonic and fermionic fluctuations were evaluated at one loop, with the exact zero modes and quasi-zero modes treated through their collective coordinates. The UV contributions from the nonzero bosonic KK modes conincides with the two-dimensional renormalization of both the constituent actions and the kink--antikink interaction exactly. In particular, the matching of the renormalization on both sides requires the running of both $g^2$ and $k$. The fermionic determinant shifts the linear separation term from $\delta\epsilon$ to $\epsilon$, while the form of the quasi-modulli integral remains intact as in the previous quantum-mechanical calculation \cite{Sheu:2024hae}.

The Lefschetz-thimble prescription determines how the real and complex bion contributions enter this integral. Their combination vanishes at the supersymmetric point, whereas the two lateral continuations give complex-conjugate results away from it. 
In the small $\delta{\epsilon}$ limit, we can further expand it and see that the imaginary ambiguity starts from the second order $E^{(2)}$ for the one-bion amplitude. This distinction is significant. The ambiguity is present within the leading one-loop bion calculation, which is valid for arbitrary $\delta{\epsilon}$, but appears only at second order in the supersymmetry-breaking deformation. This result is also consistent with the exact quantum-mechanical analysis of Ref.~\cite{Sheu:2024sxf}.

The general-target calculation makes clear which ingredients depend on the detailed metric. The moment-map difference $\Delta_{\mu,R}$ determines the constituent action, and the endpoint normalizations is involved in the final amplitude through $G_*=\sqrt{G_-G_+}$. Although $\Delta_{\mu,R}$ is expressed through endpoint values, it contains the integral of the metric between the two vacua. Local curvatures and the full profile also enter intermediate fluctuation terms and corrections beyond the leading order. The $U(1)$ weights $\pm1$, together with the twist parameter $m$, fix the linear separation term, the fermionic shift, and the logarithmic renormalization of the bion action. The resulting Stokes structure and the infrared scaling $(\Lambda R)^{4mR}$ are are shared the same form among the targets considered here. Setting $G_-=G_+=1$ and $\Delta_{\mu,R}=m\arccosh k_R/\sqrt{k_R^2-1}$ recovers the deformed $\cp^1$ result.

Picking $g_R^2$ as the coupling of the asymptotic expansion at fixed reference metric, cancellation of the one-bion ambiguity requires a positive-axis Borel singularity at $t_{\rm b}=4\pi R\Delta_{\mu,R}$. At order $(\delta\epsilon)^2$, the leading ambiguity has the form produced by the simple pole in \cref{eq:predictedborelpole}. Thus the pole position depends on the target space geometry, even though the bion weight has the same infrared scaling along the RG flow. This provides the perturbative singularity required by the semiclassical calculation. An explicit evaluation of the perturbative Borel transform around the trivial vacuum at finite radius is necessary to verify the pole and its residue, extending the explicit ambiguity cancellation found in the quantum-mechanical limit \cite{Sheu:2024hae}.

The present results apply at leading dilute one-loop order. Finite overlap corrections and higher-loop fluctuations can introduce further dependence on the metric and modify the singular behavior beyond this approximation. One can also include configurations containing several neutral bions and their interactions. Their leading actions suggest further Borel singularities at integer multiples of $t_{\rm b}$, but the corresponding coefficients and ambiguity cancellations require the multi-bion integrals. These calculations would test how the geometric structure found here extends to higher semiclassical orders.
As we restrict to the two dimensional target space with a $U(1)$ symmmetry, one may further consider a more general higher dimensional target space such as a symmetirc spaces or toric varieties.

\section*{Acknowledgments}
The author is grateful to Tatsuhiro Misumi for very useful discussions.
This work is supported by a KIAS Individual Grant QP111601 via the Quantum Universe Center at Korea Institute for Advanced Study.

{
\setlength\bibitemsep{2pt}
\printbibliography[heading=bibintoc,
title={References}]
}

\end{document}